\documentclass{ws-procs9x6}

\begin{document}

\title{Bose-Einstein Correlations as correlations of fluctuations}  

\author{O.V.Utyuzh and G.Wilk}

\address{The Andrzej So\l tan Institute for Nuclear Studies; 
Ho\.za 69; 00-689 Warsaw, Poland\\E-mail: utyuzh@fuw.edu.pl and
wilk@fuw.edu.pl}

\author{M.Rybczy\'nski and Z.W\L odarczyk}

\address{Institute of Physics, \'Swi\c{e}tokrzyska Academy;
Konopnickiej 15; 25-405 Kielce, Poland\\
E-mail: mryb@pu.kielce.pl and wlod@pu.kielce.pl}


\maketitle

\abstracts{
The limitations of the recently proposed new method of numerical
modelling of Bose-Einstein correlations (BEC) are explicitly
demonstrated. It is then argued that BEC should still be considered
as emerging from the correlations of fluctuations, however they have
to be modelled first in any Monte Carlo event generator (MCEG) and
not added {\it a posteriori} to the existing output of some MCEG. 
}

Recently we have proposed new method of numerical modelling of BEC
which makes use of the fact that BEC can be regarded as correlations
of fluctuations \cite{NEWBEC1,NEWBEC2,NEWBEC3}. Our aim was to
provide a fast algorithm, which could be used on the even-by-event
basis together with the output of other MCEG's and which would
introduce a characteristic structure of the correlation function
$C_2(Q)$ between them \cite{NEWBEC3}. The main idea was
\cite{NEWBEC1,NEWBEC3} to resign from the initial allocation of
charges to the particles provided by a given MCEG and to perform then
the new allocation in a way which would result in a desired bunching
of like particles in the phase space\footnote{At the same time all
positions of the initial particles in the phase-space, as well as the
number of particles of given charge in an event, remain intact
\cite{NEWBEC1,NEWBEC2,NEWBEC3}.}. As a result one gets a number of
(what we have called \cite{NEWBEC3}) {\it elementary emitting cells}
(EEC), $N_{cell}$, each of them containing particles, $n_{part}$, of
a given charge distributed according to geometrical distribution (in
order to model the bosonic character of produced secondaries
\cite{NEWBEC1,NEWBEC2}). Those cells (and parameters connected with
their formation) represent therefore the main object of interest in
our approach. It means then that their number and the (mean)
multiplicity of secondaries they contain are the main factors
dictated the shape of $C_2(Q)$ (and therefore also the 'size'
parameter $R$ and 'chaoticity' parameter $\lambda$ usually used to
fit $C_2(Q)$ \cite{NEWBEC1,NEWBEC3}). In Fig. 1 we show that it is,
in principle, possible to fit experimental data. Our reservation
comes from the fact that so far only one sample of $e^+e^-$ data were
fitted by only two types of MCEG (cf. \cite{NEWBEC3} for details) and
using only direct pions. But it shows also explicitly the price to
be paid for this, namely the anticorrelations between the
unlike-particles showing up\footnote{We are deeply indebted to Prof.
W.Kittel for suggesting to us this simple check before proceeding
with any further, more involved, development of our algorithm.}.

\vspace{-5mm}
\begin{figure}[h]
\begin{minipage}[t]{0.475\linewidth}
\centering
\includegraphics[height=5.cm,width=5.cm]{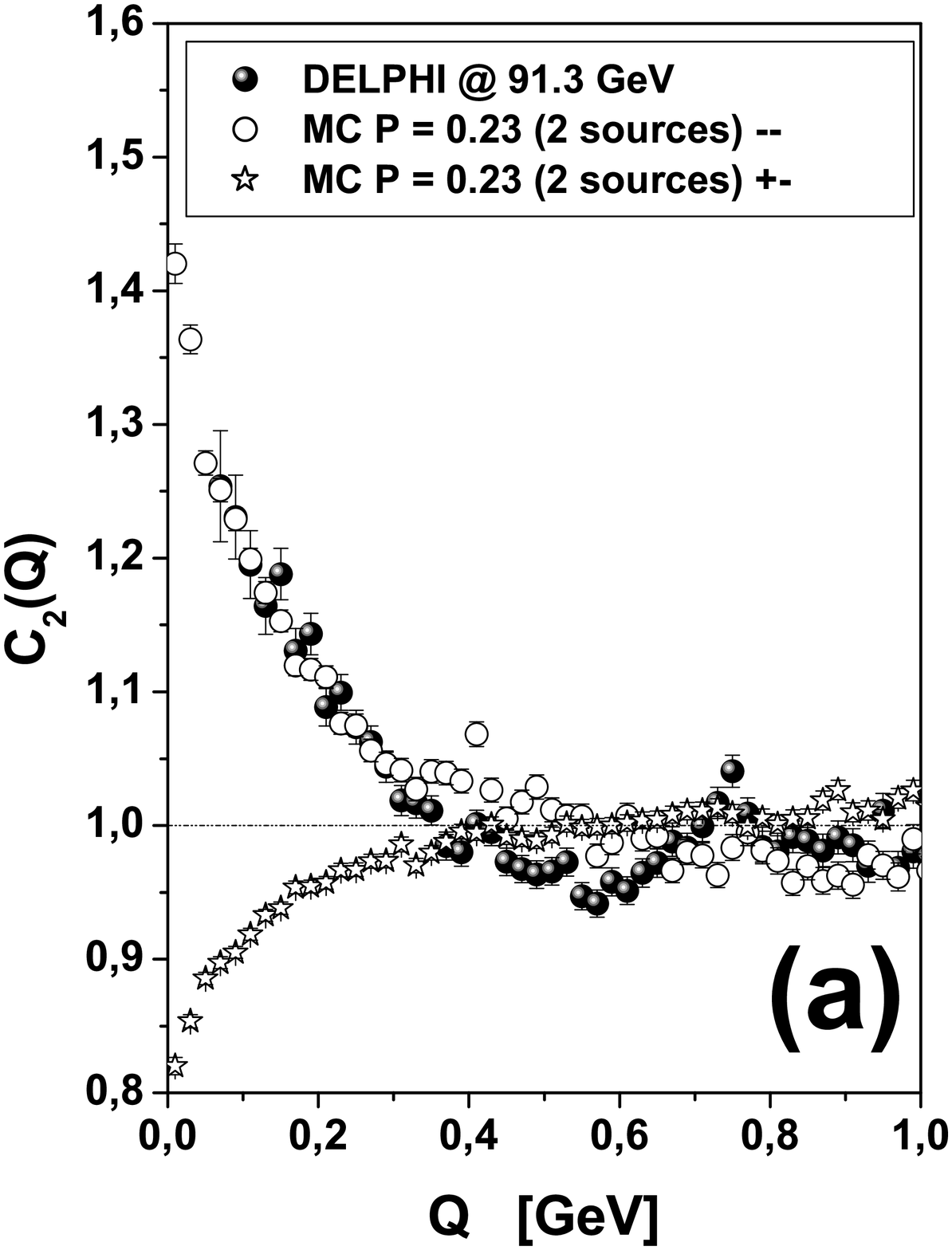}
%\vspace{-5mm}
\caption{Comparison with DELPHI data presented in
\protect\cite{NEWBEC3} shown together with the  
results for the unlike-pairs for CAS model \protect\cite{NEWBEC3}.}
\end{minipage}\hfill
\begin{minipage}[t]{0.475\linewidth}
\centering
\includegraphics[height=5.cm,width=5.cm]{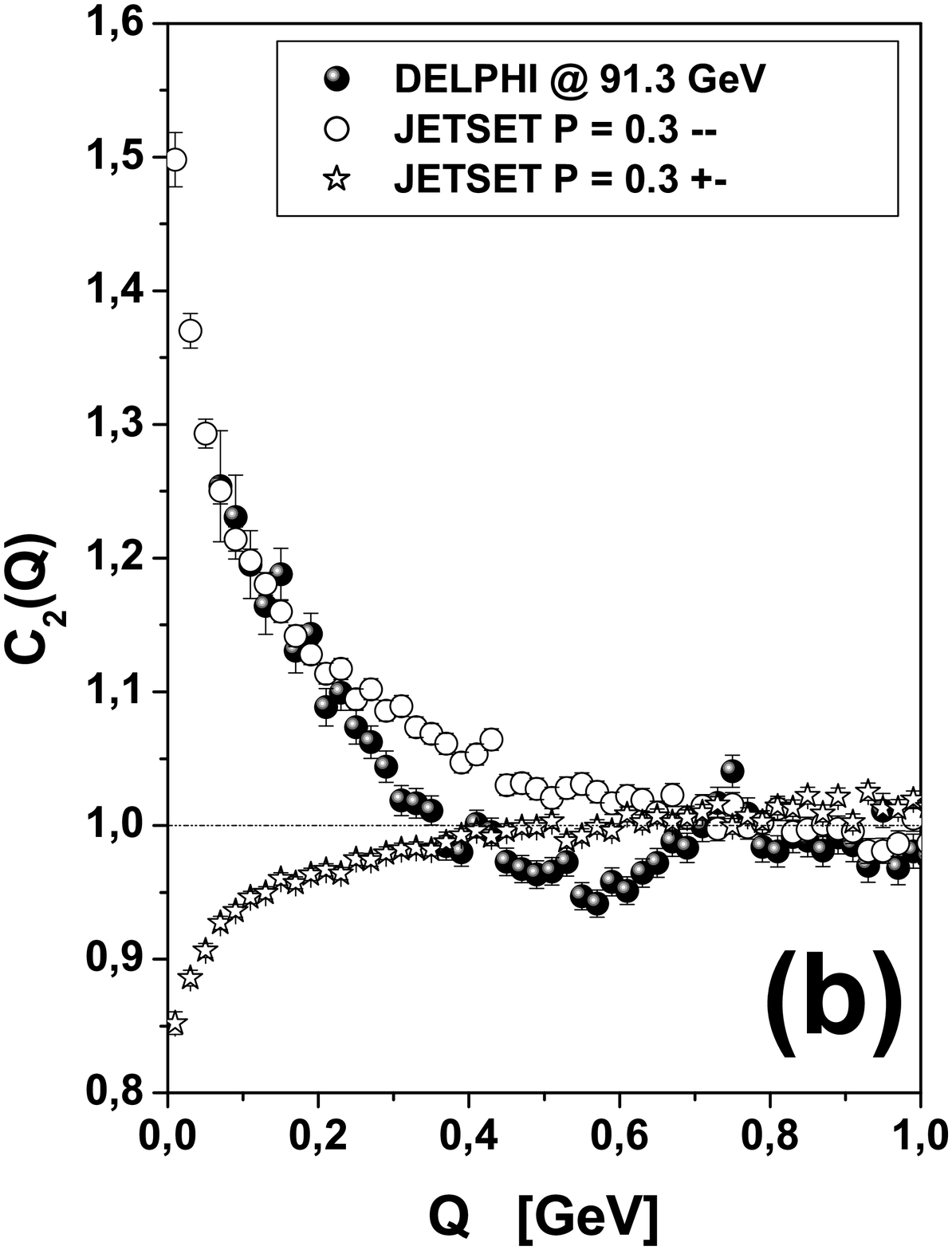}
%\vspace{-5mm}
\caption{Comparison with DELPHI data presented in
\protect\cite{NEWBEC3} are shown together with the 
results for the unlike-pairs for JETSET model
\protect\cite{NEWBEC3}.} 
\end{minipage}
\end{figure}
Such anticorrelation feature is, however, after a bit of thinking, to
be expected. The reason is that we start, in a given event, with {\it
a priori} uncorrelated particles provided by a MCEG. Our algorithm
shifts now charges in a fixed pool of particles in this event. It is
constructed in such a way that the EEC of different charges
essentially do not overlap in the phase space (although, in principle
there is nothing preventing them to do so; but we have checked that
allowing for substantial overlap in order to diminish the
anticorrelation effect weakness dramatically the obtained
BEC signal as well). The possible way out is the reparametrization of
the original MCEG in such way as to have {\it positive}
charge-particles correlations from the very beginning which are then
washed out by the action of our algorithm. 
%\vspace{-1cm}
\begin{figure}[h]
\begin{center}
\includegraphics[height=12.cm,width=11.cm]{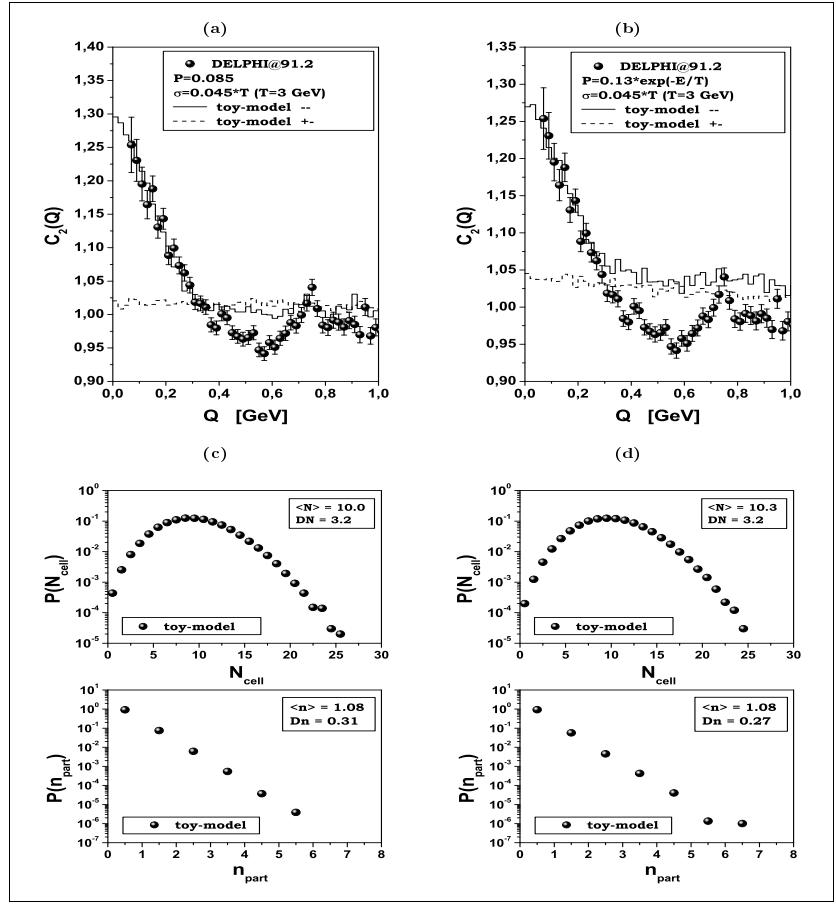}
\end{center}
\caption{$(a,b)$ Application of our toy model with two types of
weights $P$ (defining the formation of the EEC's) to fit DELPHI data
\protect\cite{NEWBEC3} on $e^+e^-$ annihilation. Panels $(c)$ and
$(d)$ show the corresponding distributions of EEC, $P(N_{cell})$, and
distribution of the like-charge particles allocated to such cell,
$P(n_{part})$, with mean values and dispersions as indicated
(corresponding also to parameters of the best fits, respectively,
poissonian and geometrical, to $P(N_{cell})$ and $P(n_{part})$, which
are not shown here). Notice the small values of multiplicities in
EEC's, $\langle n\rangle \sim 1.1$ only.}   
\end{figure}
Another, more physical (in our opinion) possibility, which we shall
demonstrate here, is to construct MCEG satisfying the requirements
of BEC from the very beginning (i.e., essentially to continue the
line of reasoning proposed some time ago in \cite{OMT}). It means
that, contrary to the situation in Figs. 1 and 2, where the original
outputs of MCEG were not showing any BEC and our algorithm was not
changing the main characteristics of theses outputs (like, for
example, multiplicity distributions), one has to change from the very
beginning the inclusive distributions to accommodate the
characteristic BEC effects in them, see also discussion in
\cite{NEWBEC3}). Fig. 3 shows results of our first attempts in this
direction \cite{NEWBEC3} where preliminary comparison of a kind of
toy model (presented in \cite{NEWBEC3}) with data at $W=91.2$ GeV has
been shown for two types of the specific weight variables, $P$
\cite{NEWBEC3}.  Particles were selected from the energy $W$ with
given proportion of $(+/-/0)$ and allocated to EEC's in such 
a way that, as evident in Fig. 3, distribution of cells, $P(N_{cell})$,
is poissonian whereas distribution of particles (of the same sign) in
a cell, $P(n_{part})$, is geometrical. It means then that the
multiplicity distribution of all particles follow the Negative
Binomial (NB) form with EEC playing a role of {\it clans}
\cite{NEWBEC3}. It is very important to realize that in order to get
the observed structure of $C_2(Q)$ one has to allow for some
uncertainties in energies of particles allocated to a given EEC (here
given by a gaussian form with width $\sigma$ as indicated in Fig. 3,
they correspond to the {\it sizes} of EEC, $\delta y$, introduced in
\cite{OMT}, although they are not identical with them). As one can
see the result of our "toy model" shown in Fig. 3 is quite promising
and can form a basis of more detailed investigations (which we plan
to pursue\footnote{In particular one should improve the approximate
for a moment (although kept on the same numerical level as in
\cite{OMT}) conservation of charges and energy-momenta in a given
event.}).  

\section*{Acknowledgments}
One of us (GW) would like to thank the Bogolyubow-Infeld Program
(JINR, Dubna) for financial help in attending the XXXII ISMD
conference where the above material has been presented.

\end{document}